\documentstyle[psfig]{mn}

\newif\ifAMStwofonts
\AMStwofontstrue

\title[Red Clump distance to the LMC]
{A theoretical analysis of the 
systematic errors in the Red Clump distance to the LMC}
\author[M. Salaris et al.]
       {Maurizio Salaris$^{1}$, Susan Percival$^{1}$ and L\'eo Girardi$^{2}$\\
$^1$Astrophysics Research Institute, Liverpool John Moores
        University, Twelve Quays House, Birkenhead CH41 1LD, 
	UK \\
$^2$INAF -- Osservatorio Astronomico di Trieste, 
    Via Tiepolo 11, I-34131 Trieste, Italy}
\date{Accepted 2003 ???.
      Received 2003 ???;
      in original form 2003 ???}

\pagerange{\pageref{firstpage}--\pageref{lastpage}}
\pubyear{2003}

\begin{document}

\maketitle

\label{firstpage}

\begin{abstract}

We present a detailed analysis of the uncertainty on the theoretical
population corrections to the LMC Red Clump (RC) absolute magnitude,
by employing a
population synthesis algorithm to simulate theoretically 
the photometric and spectroscopic properties 
of RC stars, under various assumptions about the
LMC Star Formation Rate (SFR) and Age Metallicity Relationship (AMR).
A comparison of the outcome of our simulations with 
observations of evolved
low-intermediate mass stars in the LMC allows one to
select the combinations of SFR and AMR 
that bracket the real LMC star formation history, 
and to estimate the systematic error on the 
associated RC population corrections.

The most accurate estimate of the LMC distance modulus from the RC
method (adopting the OGLE-II reddening maps for the LMC)
is obtained from the $K$-band magnitude, and provides
$(m-M)_{0, LMC}=18.47\pm0.01(random)^{+0.05}_{-0.06}(systematic)$.
Distances obtained from the $I$-band, or from the multicolour
RC technique which determines at the same time reddening and distance,
both agree (albeit with a slightly larger error bar) with this value.
 
\end{abstract}


\section{Introduction}
\label{sec_intro}

The determination of the Star Formation Rate (SFR) and  
Age-Metallicity-Relationship (AMR) of the stellar populations in a
generic galaxy is a fundamental step in order to understand 
galactic formation mechanisms. In addition to this, a reliable
determination of the SFR and AMR in galaxies with a resolved stellar
population is necessary if their distance is to be determined by means
of the Red Clump (RC) method (Paczynski \& Stanek~1998).

The relevance of RC stars (helium burning 
stars where helium ignited in an electron degenerate core,
plus helium burning stars of higher masses but of similar
absolute magnitude) as distance indicators stems from the fact that
$Hipparcos$ parallaxes allow an extremely accurate calibration of the average
RC brightness in the solar neighbourhood (with an uncertainty of only
$\pm$0.03 mag, see e.g. Stanek \& Garnavich~1998; 
Alves et al.~2002), and that the RC is easily
recognizable in the Colour-Magnitude-Diagram (CMD) of intermediate-old
stellar populations.
The determination of the absolute magnitude of the local RC in a given
passband $\lambda$,
$M_{\lambda, local}^{\rm RC}$, and 
the apparent magnitude $m_\lambda^{\rm RC}$ of the
RC in a given stellar population is not difficult,
since in both the {\em Hipparcos} database of nearby stars,
and in CMDs covering even a small fraction of a nearby galaxy, 
one finds several hundreds of clump stars, easily 
identifiable from their CMD location. 
As proposed by 
Stanek \& Garnavich~(1998), a non-linear least-square fit of the function 
\begin{equation}
N(m_\lambda) = a + b m_\lambda + c m_\lambda^2 + 
	d \exp\left[-\frac{(m_\lambda^{\rm RC}-m_\lambda)^2}{2\sigma_{m_\lambda}^2}\right]
\label{eq_fit}
\end{equation}
to the histogram of stars in the clump region per magnitude bin 
provides the value of 
$m_\lambda^{\rm RC}$ and its associated standard error.

The local RC absolute brightness is known with high accuracy,
whereas the dependence of the RC brightness 
on the SFR and AMR of a generic stellar population has been
the subject of many papers published in the last 5 years. Cole~(1998),
Girardi et al.~(1998), Sarajedini~(1999) and, 
in great detail, Girardi \& Salaris~(2001,
Paper~I), Salaris \& Girardi~(2002, Paper~II), Percival \&
Salaris~(2003, Paper~III)
have conclusively demonstrated that the brightness of
RC stars in a given passband shows a non-monotonic, complicated
dependence on age and metallicity. On average, the $V$- and $I$-band
magnitude of RC stars are more sensitive to [Fe/H] and age than the
$K$-band, but the reverse is true for high ages and low metallicities.
Differences up to $\sim$0.3 mag with respect to the local RC are
possible for particular combinations of age and metallicity.

The approach we followed in Papers~I, II and III 
to study the age and metallicity dependence of the RC brightness
has been based on the use of stellar models (Girardi et al.~2000)
in a purely differential way. By means of a population synthesis
algorithm we have computed a syntethic CMD 
of the local RC by using the SFR and AMR by Rocha-Pinto et
al.~(2000a,b -- see detailed discussions in Paper~I, II and III) and
derived $M_{\lambda, local}^{\rm RC, theory}$ from fitting
eq.~\ref{eq_fit} to our data.
Once the SFR and AMR of another observed stellar population are specified, 
they are employed to compute the appropriate CMD and derive 
$M_{\lambda, galaxy}^{\rm RC, theory}$. 
The difference 
$\Delta M_\lambda^{\rm RC} = M_{\lambda, local}^{\rm RC, theory} - 
M_{\lambda, galaxy}^{\rm RC, theory}$ we call the population correction.
To determine the absolute magnitude of the RC in the selected
population one has to
subtract this theoretically determined $\Delta M_\lambda^{\rm RC}$
from the observed absolute brightness of the local RC, $M_{\lambda, local}^{\rm
RC}$, determined from $Hipparcos$ data.
As for the distance,
once the mean apparent magnitude of the RC in a given photometric
band, $m_{\lambda}^{\rm RC}$, 
is measured in the population under scrutiny, its distance modulus 
is easily derived by means of
$(m-M)_0 = m_\lambda^{\rm RC} - M_{\lambda, local}^{\rm RC} 
- A_{m_\lambda} + \Delta M_\lambda^{\rm RC}$,
$A_{m_\lambda}$ being the interstellar extinction.

In Paper~III we used a sample of Galactic open clusters with
a range of [Fe/H] and ages in order to test empirically the accuracy of our 
$\Delta M_\lambda^{\rm RC}$ determinations. The distances to the
clusters have been determined from a purely empirical 
Main Sequence-fitting technique,
by using a large sample of field dwarfs with accurate $Hipparcos$
parallaxes and [Fe/H] values; 
in this way absolute magnitudes $M_{\lambda, cluster}^{\rm RC}$
of their RC stars have
been determined and reliably tied to $Hipparcos$ parallaxes. The
empirical differences  
$M_{\lambda, local}^{\rm RC} - M_{\lambda, cluster}^{\rm RC}$
have been then computed and compared with our predicted 
$\Delta M_\lambda^{\rm RC}$ on a cluster by cluster basis. 
Our analysis has shown
that our theoretical population corrections do reproduce accurately their
empirical counterparts, with no statistically significant offset, nor trend
with respect to either [Fe/H] or age.
Since these $\Delta M_\lambda^{\rm RC}$ values for single-age, 
single-metallicity systems are the building blocks for computing corrections 
in the case of a more complicated SFR and AMR, the accuracy of the population
corrections to external galaxies will therefore depend entirely 
on how accurately their SFR and AMR have been determined.

In this paper we will discuss in detail the case of the Large
Magellanic Cloud (LMC). The LMC plays a fundamental role in
establishing the extragalactic distance scale, since the zero point of
the Cepheid Period-Luminosity (P-L) relationship -- the cornerstone of the
cosmological distance ladder -- is set by the LMC distance.
The value of the Hubble constant determined by the $HST$ extragalactic
distance scale key project (Freedman et al.~2001) is based on a zero
point for the Cepheid P-L relationship obtained assuming a 
LMC distance modulus equal to $(m-M)_{0}=18.50\pm$0.10.

The LMC is probably also the external galaxy for which there is the
largest number of data from which one can derive the SFR and AMR.
In Paper~I and II we provided the values of $\Delta M_\lambda^{\rm
RC}$ for the LMC in the Johnson $V$-, Cousins $I$- and 
Bessel \& Brett $K$-band, computed using recent
determinations of the LMC star formation history. Our population
corrections coupled to the local RC calibration provide a LMC distance
modulus $\sim$18.50 (Alves et al.~2002; Paper~II; Paper~III).
This distance strongly supports the Cepheids P-L zero point
assumed by the $HST$ key project.

The question we wish to address here for the first time is: 
what is a realistic systematic uncertainty 
associated with our $\Delta M_\lambda^{\rm RC}$ correction for the LMC ?
Some estimates of this uncertainty have appeared in the
literature (e.g., $\pm$ 0.03 mag for the $K$-band, according to Alves
et al.~2002, or a more pessimistic general evaluation of $\pm$0.15 mag 
by Pietrzy\'nski, Gieren \& Udalski~2003), but a rigorous analysis of
this issue is still lacking.

To address this problem we will produce theoretical CMDs for the 
LMC field stellar population, by spanning a range
of proposed SFRs and AMRs. The results of our simulations 
will be compared with photometric data of RC stars
and metallicity determinations of red giants belonging to the inner
regions of the LMC;
the aim is to determine the possible range of $\Delta
M_\lambda^{\rm RC}$ allowed by observations of the evolved
intermediate-old population. It is very probable that some of
the AMR and SFR combinations allowed by the evolved stars can be ruled
out by constraints imposed by earlier evolutionary phases; in this case, our
determination of the systematic error on the RC luminosity will be a 
safe upper limit to the real uncertainty. 

In Sect.~\ref{sec_sim} we discuss our population synthesis computations
for the various SFR and AMR assumed;
in Sect.~\ref{sec_confphot} we test the results from our simulations
against RC photometric constraints, 
whereas in Sect.~\ref{sec_confmet} we study the constraints posed by
metallicity estimates of red giant stars.
Our final results and conclusions are presented in Sect.~\ref{sec_concl}. 

\section{CMD simulations and results}
\label{sec_sim}

As in our previous paper we will base the discussion of the model 
behaviour on the Girardi et al.~(2000) set of
evolutionary tracks and isochrones. 
Different sets of models in the literature
present systematic luminosity differences for 
the core helium burning stars that have passed through
the helium flash (ages larger than $\sim$0.5 Gyr)
-- which are the stars belonging to the observed RC
populations -- due mainly to the different values of the  
helium core mass at the flash (see, e.g., Salaris, Cassisi \& Weiss~2002);
however, the variation of their brightness with respect to age and
metallicity is much more consistently predicted
by theory (Castellani et al.\ 2000; Salaris et al.\ 2002).
The main results of this paper will be 
based on the strictly differential use of the model predictions. 

\begin{figure}
\psfig{file=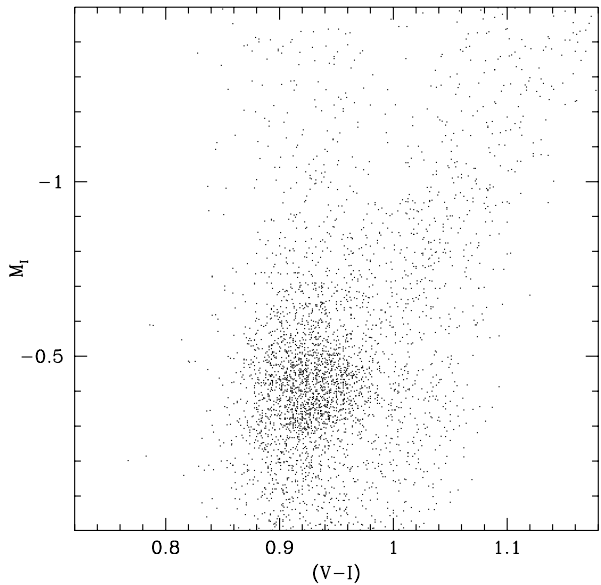,width=8.3cm}
\caption{$M_I-(V-I)$ synthetic CMD for our LMC reference model (see
text for details).}
\label{RCcmd}
\end{figure} 

Theoretical CMDs have been computed by interpolating among Girardi et
al.~(2000) models according to a specified SFR and AMR, assuming a
Salpeter Initial Mass Function (IMF) with exponent $-$2.3. We have
tested the effect of varying the IMF exponent by $\pm$0.5 around this
`canonical' value, and we found the the resulting RC brightness in
$V$, $I$ and $K$ is affected at most at the level of 0.01 mag (in the
case of the $V$-band). 
In all our synthetic CMDs we have
included a 1$\sigma$ gaussian photometric error of 0.02 mag to simulate typical
photometric errors (the exact value of this quantity 
does not affect at all the results of
our investigation), as well as a given
spread around the mean AMR. 
The RC of the synthetic CMDs is populated typically by
$\sim$1500 objects.

All the simulations discussed in this paper -- 
which are summarized in Table~1 -- produce a RC that, with
the inclusion of the photometric error, well reproduces the 
approximately round and featureless shape of the observed 
$I-(V-I)$ CMDs of the LMC (see Fig.~\ref{RCcmd} for the result of 
a typical simulation);
in particular, they resemble the many CMDs presented by the OGLE
group (e.g., Udalski et al.~2000).\footnote{We remark that, under 
excellent photometric conditions and with homogeneous reddening, 
the observed LMC clump may present substructures 
which are predicted by theory, as discussed in Girardi (1999).} 

\begin{figure}
\psfig{file=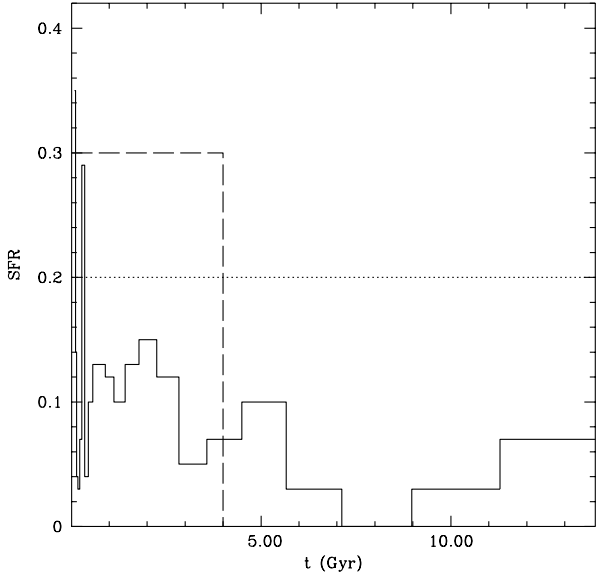,width=8.3cm}
\caption{SFRs employed in our simulations. The solid line displays the
Holtzman et al.~(1999) SFR we use in our reference model, while the dotted
and dashed lines show the other two extreme cases we use in our
simulations (see text for details). The vertical axis is in
arbitrary units of $M_{\odot} yr^{-1}$.}
\label{SFR}
\end{figure}

\begin{figure}
\psfig{file=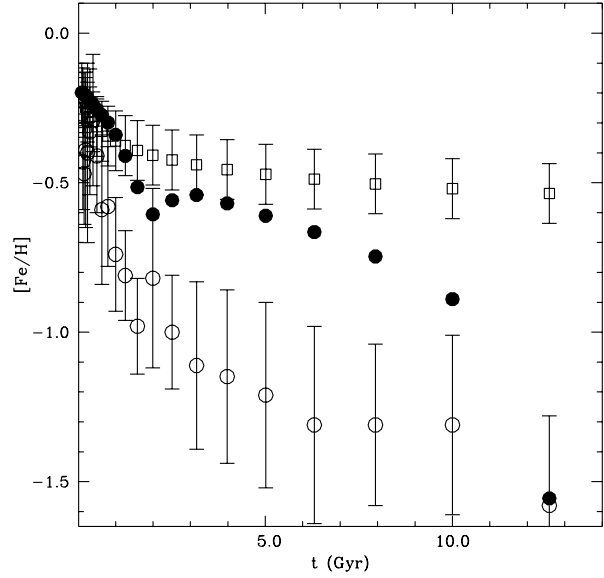,width=8.3cm}
\caption{AMRs employed in our simulations. Filled circles represent the
Pagel \& Tautvaisiene~(1998) one, while open squares and open circles
display, respectively, the Dopita et al.~(1997) AMR and the Dirsch et
al.~(2000) ones.}
\label{AMR}
\end{figure}

We have considered a set of SFRs and AMRs for the LMC, as displayed in
Figs.~\ref{SFR} and ~\ref{AMR}, which cover the actual range
of empirical determinations. The SFR by Holtzman et al.~(1999 - H99)
displayed in Fig.~\ref{SFR} as a solid line, 
has been obtained from deep CMDs 
(including Main Sequence stars) of the bar fields, coupled to the 
theoretical AMR by Pagel \& Tautvaisiene~(1998 -- PT98)
shown as filled circles in Fig.~\ref{AMR}.  This combination of SFR and
AMR constitutes our `reference' model.
There is general consensus about the fact that the SFR of the LMC
field had a marked increase around 2-4 Gyr ago (see, e.g., the
discussion by Dirsch et al.~2000 and references therein), as 
prescribed by our reference SFR. 
However, the exact amount of this increase is somewhat 
subject to uncertainty; as an extreme case we therefore 
assumed that the increase
has been negligible, hence the constant SFR displayed as a dotted line
in Fig.~\ref{SFR}, which approximates well (for the age range relevant
to RC stars) the SFR determined by
Smecker-Hane et al.~(2002) for a disk LMC field. 
The opposite extreme case is to assume a negligible (i.e. zero) SFR
until $2-4$ Gyr ago, and then a constant one; we choose therefore 
a SFR constant between now and 4 Gyr ago, and equal to zero before
(dashed line in Fig.~\ref{SFR}). Moving the upper age limit from 4 Gyr
down to 2.5 Gyr does not influence the results about the evolved stars in
our simulations, for the reasons discussed later on.
It is important to remark that a SFR larger in the first Gyrs of galaxy
evolution can be ruled out by the observed shape of the RC in the
$I-(V-I)$ CMD. 
This is the case, for example, for the SFR by 
Dolphin~(2000), which is
characterized by a formation rate higher in the first 4 Gyr of the galaxy
evolution, than between 0.5 and 2.5 Gyr ago; as discussed in
Paper~I -- and noticed also by Dolphin~(2000) --  
the resulting RC morphology is at odds with the
observations.
In fact, it shows a substantial tail of stars bluer than the
main body of the RC, which are not observed (this tail is particularly
pronounced in the $K$-band, and it is not observed in this photometric
band either).
This result is independent of the AMR used; in fact, we obtain the 
same blue tail when using either the Dolphin~(2000) AMR or the 
PT98 one (the AMR provided by
Dolphin is similar to PT98 results for ages less than $\sim 2-3$ Gyr 
whereas it approaches the more metal poor AMR by 
Dirsch et al.~2000 at higher ages --
see below for more details about the LMC AMR ).

We have also tested the recent SFR determined by Smecker-Hane et
al.~(2002) for a bar field, obtained assuming essentially the PT98
AMR. This SFR is not very different from the H99 one and we have
verified that it provides the same RC brightness (as well as mean [Fe/H]
and age) as the H99 SFR when the same AMR is used.

As for the AMR, we have employed the already mentioned PT98 one, plus
two other determinations found in the literature.
The first one is from Dirsch et al.~(2000 -- DI00), displayed as open
circles in  Fig.~\ref{AMR}; it is based on
photometric metallicity determinations of samples of field red giant stars
(from Str\"omgren photometry), coupled with ages determined from
isochrone fitting, and predicts much lower metallicities at a given
age than the PT98 results. 
The second one is the AMR by Dopita et al.~(1997 -- DP97), shown as
open squares in Fig.~\ref{AMR}, which is based on spectroscopy
of planetary nebulae, and modelling of the hot central stars to
determine their age;
this AMR appears to be slightly more metal rich on average than the
PT98 one\footnote{DP97 determine abundances of $\alpha$-elements 
which we transformed into Fe abundances assuming a scaled solar metal 
distribution; see e.g. the discussion in DI00.}. 
We have also checked 
the AMR proposed by Olsen~(1999), based on the AMR of LMC star
clusters of various ages. This AMR predicts 
metallicities that are intermediate between PT98 and DP97 
for the age range relevant to RC stars, and the results of the simulations behave 
accordingly.

Starting from these choices of AMR and SFR we have computed 3 sets of
galaxy population models, each set made 
up of 3 models with the same choice of AMR, and for the 3
alternative SFRs; all computed models are summarized in Table~1.
Set I has been calculated by employing the PT98 AMR,
set II has made use of the DI00 AMR, and set III the DP97 one.
The suffix $a$, $b$ or $c$ denotes the choice of the SFR; $a$ corresponds to
the H99 SFR, $b$ to a constant one, $c$ to a SFR constant for the last
4 years of galaxy evolution and zero before.

The `reference' model is model I$a$, computed with 
our standard assumption about SFR and AMR, that is  -- as in our
previous papers -- the SFR by H99
and the AMR by PT98; we have 
added a Gaussian spread by 0.10 dex to this AMR, however the precise
value of this spread does not affect appreciably the results.

\begin{figure}
\psfig{file=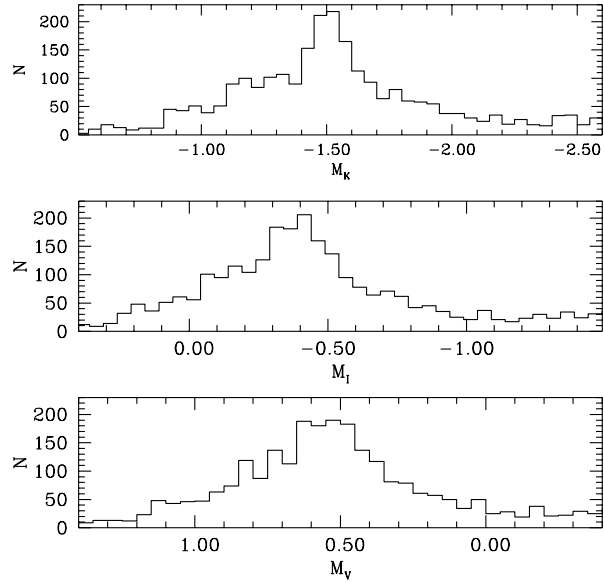,width=8.3cm}
\caption{Number of stars per magnitude bin in the RC region of the CMD
corresponding to our reference model.}
\label{Hist}
\end{figure}

Figure~\ref{Hist} shows, as an example, the histogram of the 
number of stars per magnitude bin in the RC region for this reference
model I$a$ of the LMC population. By fitting equation~\ref{eq_fit}
to these histograms we determine $M_{\lambda, LMC}^{\rm RC, theory}$
in $V$, $I$ and $K$; by using the value of
$M_{\lambda, local}^{\rm RC, theory}$ from our simulations of the solar
neighbourhood ($M_{K, local}^{RC, theory}=-1.54$,
$M_{I, local}^{RC, theory}=-0.17$, $M_{V, local}^{RC, theory}=0.84$)
we then computed $\Delta M_\lambda^{\rm RC}$. Typical errors on  
$\Delta M_\lambda^{\rm RC}$ as obtained from the gaussian fitting to
the RC number counts are equal to $\sim$0.01 mag.
The full results for our ensemble of simulations are summarized in Table~1.

\begin{figure}
\psfig{file=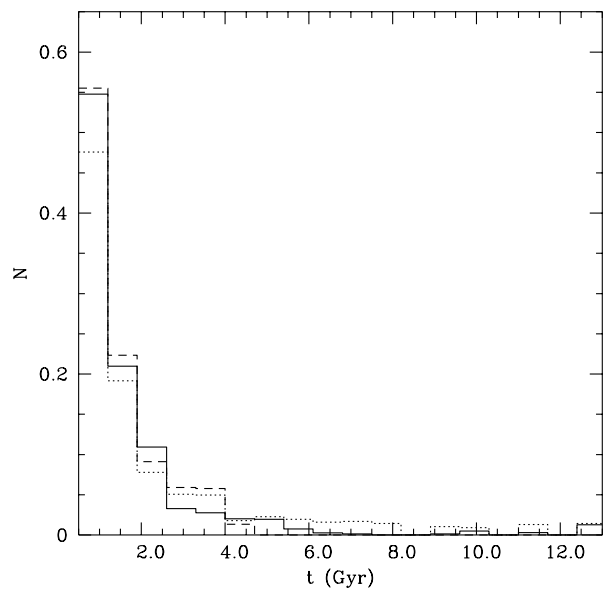,width=8.3cm}
\caption{Histogram of the number of RC stars as a function of age 
in our reference model~I$a$ (solid line), model~I$b$ (dotted line)
and model~I$c$ (dashed line). The total number of RC stars in each
model has been normalized to 1.}
\label{Histage}
\end{figure}

\begin{figure}
\psfig{file=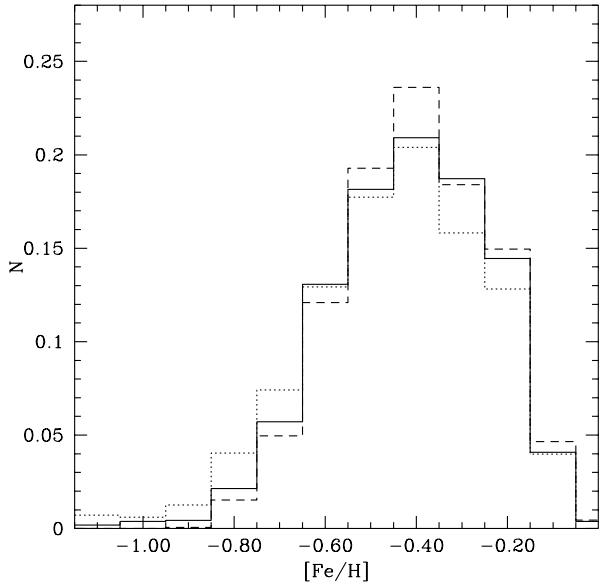,width=8.3cm}
\caption{Histogram of the number of RC stars as a function of [Fe/H] 
in our reference model~I$a$ (solid line), model~I$b$ (dotted line)
and model~I$c$ (dashed line). The total number of 
RC stars in each model has been normalized to 1.}
\label{Histmet}
\end{figure}

The distributions of age and metallicity 
of RC stars are displayed in Figs.~\ref{Histage} 
and \ref{Histmet}. It is evident, from a comparison with 
Fig.~\ref{SFR}, that the age distribution does not
follow exactly the SFR; this occurrence has already been discussed 
in detail in 
Paper~I and stems from the fact that for RC stars the ratio between 
Main Sequence
and helium burning timescales is not constant, but decreases for
increasing stellar mass (hence decreasing age). This explains the
large increase in the number of RC stars with ages lower than about 
$2.0-2.5$~Gyr, which is larger than what is expected on the basis 
of the adopted SFR alone. The [Fe/H] distribution is characterized by a mean
metallicity equal to [Fe/H]$\sim -$0.40 which corresponds, 
according to the adopted AMR, to a mean age of about 1.6 Gyr. 


\begin{table*}
\caption{Absolute $V$, $I$ and $K$ magnitude of the RC in the solar
neighbourhood (as determined from $Hipparcos$ parallaxes) and 
the population corrections for the LMC
as determined under various assumptions about the SFR and AMR.}
\label{tab_corrections}
\begin{tabular}{lrrrll}
\noalign{\smallskip}\hline\noalign{\smallskip}
Model & $M_{V}^{\rm RC}$ & $M_{I}^{\rm RC}$ &
$M_{K}^{\rm RC}$ & & \\
\noalign{\smallskip}\hline\noalign{\smallskip}
Solar neighbourhood & 0.73$\pm$0.03 & $-0.26\pm$0.03 & $-1.60\pm$0.03
& (from $Hipparcos$ parallaxes&) \\
\noalign{\smallskip}\hline\noalign{\smallskip}
  & $\Delta M_{V}^{\rm RC}$ & $\Delta M_{I}^{\rm RC}$ &
$\Delta M_{K}^{\rm RC}$ & SFR & AMR \\
\noalign{\smallskip}\hline\noalign{\smallskip}
I$a$ (reference)&0.26$\pm$0.01 & $0.20\pm$0.01 & $-0.03\pm$0.01& H99 & PT98\\
I$b$ & 0.23$\pm$0.01 & $0.16\pm$0.01 & $-0.03\pm$0.01&constant& PT98\\
I$c$ & 0.27$\pm$0.01 & $0.21\pm$0.01 & $-0.02\pm$0.01&constant t$<$4Gyr&PT98\\
\noalign{\smallskip}\hline\noalign{\smallskip}
II$a$ & 0.42$\pm$0.01 & $0.29\pm$0.01 & $-0.07\pm$0.01& H99& DI00\\
II$b$ & 0.40$\pm$0.01 & $0.27\pm$0.01 & $-0.07\pm$0.01&constant&DI00\\
II$c$ & 0.41$\pm$0.01 & $0.28\pm$0.01 & $-0.07\pm$0.01&constant t$<$4 Gyr&DI00\\
\noalign{\smallskip}\hline\noalign{\smallskip}
III$a$ & 0.20$\pm$0.01 & $0.17\pm$0.01 & $-0.03\pm$0.01& H99& DP97\\
III$b$ & 0.15$\pm$0.01 & $0.12\pm$0.01 & $-0.03\pm$0.01&constant&DP97\\
III$c$ & 0.16$\pm$0.01 & $0.16\pm$0.01 & $-0.02\pm$0.01&constant t$<$4 Gyr&DP97\\
\noalign{\smallskip}\hline\noalign{\smallskip}
\end{tabular}
\end{table*}

It is interesting to notice in Figs.~\ref{Histage} and 
\ref{Histmet} that, in the case of a model with the same AMR but
constant SFR (model I$b$), the age (mean age of RC stars equal to
$\sim$2.0 Gyr in this model) and [Fe/H] distribution of RC
stars is not altered dramatically with respect to the reference model. 
The contribution of
the older (and more metal poor) generations of stars is still
extremely small, 
due to the already mentioned trend with mass of
the ratio between helium burning and Main Sequence timescales.
The RC [Fe/H] distribution displays in model~I$b$ a mean
value [Fe/H]$\sim -$0.44, less than 0.05 dex lower than in
case of our reference SFR, and a similar shape. 
The corresponding values 
of $\Delta M_{\lambda, LMC}^{\rm RC}$ reported in Table~1 show
an overall variation of less than 0.05 mag; the precise amount 
of this variation and its sign depend critically on the exact
distribution of age and [Fe/H] for the RC stars, due to the complex and
non monotonic behaviour of the clump brightness as a function of age
and [Fe/H] (see the in depth discussions in Paper~I, II and III), and
can't simply be estimated by comparing the average [Fe/H] and average
age of RC objects.

With model I$c$ we can explore the opposite extreme 
case in which the SFR at
ages up to 4.0 Gyr ago is constant, whereas it is equal to zero for older ages.
It is evident from Table~1
that the change in the RC brightness is again very small, of the order
of a few hundredths of magnitude; the average [Fe/H]
of the clump stars is equal to $-$0.39 dex, basically coincident with
the reference value, the average age is $\sim$1.4 Gyr (see 
Figs.~\ref{Histage} and ~\ref{Histmet}).
Similar results concerning the effect of the SFR choice 
are obtained when
considering the other sets of models, II and III.

We can therefore conclude that for all cases where the SFR is not 
decreasing from the start of the galaxy formation,
the precise shape of the SFR is not a very critical factor in determining
the RC properties; moreover, the relevant age range of RC stars is
between $\sim$0.5 Gyr and 2--3 Gyr (see also discussion in Paper~I). 
For a given choice of the AMR, even enhancing the existing uncertainties 
about the SFR of the LMC, the associated variation of the RC brightness 
is within $\sim$0.05 mag -- in fact, in the $K$-band this variation is 
no more than 0.01 mag.



We turn now to the effect of the AMR choice.
The PT98 AMR employed in models I$a$,$b$ and $c$ comes from theoretical  
computations of the LMC chemical evolution
which reproduce the average metallicity distribution of LMC clusters.
In models II$a$,$b$ and $c$ we have used instead
the more empirical AMR of DI00.
The corresponding results presented in Table~1 show a consistent 
variation of the RC brightness in the $V$- ($\approx$0.15 mag brighter),
$I$- ($\approx$0.10 mag brighter) and $K$-band ($\approx$0.04 mag
fainter) with respect to the corresponding models in set~I. 
These variations are due to the much
lower metallicities (difference of the order of 0.3 dex) 
predicted by the adopted AMR for the age range of RC stars, which is
still between $\sim$0.5 and $\sim 2-3$ Gyr, since the RC 
age distribution is not appreciably changed with respect to the results
of set~I. As discussed in Paper~II, an overall lower [Fe/H] 
makes RC stars brighter in $V$ and $I$, but fainter in $K$.

The simulations for our set~III adopt the AMR by DP97.
This AMR appears to be slightly more metal rich on average than the
PT98 one. The results for the RC brightness behave accordingly,
with the $V$ and $I$ magnitudes fainter on average by $\approx$0.1 and
$\approx$0.05 mag than the models in set~I, 
whereas the $K$ magnitude is practically unaffected.

This comparison highlights the important role played 
by the AMR. In fact, these different estimates of the LMC 
AMR can produce sizable variations of the RC brightness, 
at least in the $V$ and $I$
photometric bands; the $K$-band absolute magnitude is much less sensitive even
to these large [Fe/H] variations. The present uncertainty on the AMR
of the LMC stellar population provides therefore the largest
contribution to the uncertainty on the RC population corrections for
this galaxy in the $V$ and $I$ photometric bands.




\section{Comparison with RC photometry}
\label{sec_confphot}

A first empirical check for the RC population corrections 
set by the evolved intermediate-low mass stars 
is the observed value of the difference between the RC brightness in 
different photometric bands; if the sensitivity of the RC 
absolute magnitude to the input SFR and AMR is very different in the
two bandpasses, this kind of comparison tests the
adequacy of the adopted population model. 
We have considered the RC magnitude difference
between $I$ and $K$ -- hereafter $D_M$ -- based on the 
Pietrzy\'nski \& Gieren~(2002) $K$-band observations of OGLE-II fields 
I and II (which we transformed to the Bessel \& Brett system by
adding 0.044 mag to their published value, e.g. Pietrzy\'nski \& Gieren~2002 
and Alves et al.~2002), 
and the corresponding $I$-band data from Udalski et
al.~(1999); the adopted extinction comes from the reddening maps by
Udalski et al.~(1999), coupled with the extinction law by Schlegel,
Finkbeiner \& Davis~(1998). 
The reddening maps are based on the hypothesis of homogeneous
stellar populations in all OGLE-II fields (which basically cover the
LMC bar and parts of the inner disk), so that the RC absolute magnitude
(the $I$-band is used) is assumed constant, and
brightness differences are only due to reddening differences. The reddening
zero point is set by a series of independent determinations based on
other reddening indicators, and has an estimated uncertainty of 0.02
mag (a recent analysis by Tammann, Sandage \&
Reindl~2003 confirms the adopted reddening zero-point within this 
formal uncertainty).
The hypothesis of homogeneous RC populations 
-- at least from the point of view of the chemical abundances that,
for the range of possible SFRs for the LMC, 
are the main factor in determining the 
RC brightness -- in the bar and inner disk
is supported by the results from Bica et al.~(1998), and will be used in the
rest of the paper, where data from the inner disk and bar fields will be
treated as if belonging to the same stellar population.
It is also important to stress that
differences are found when outer LMC fields are considered (e.g. the
discussions in Bica et al.~1998; Olsen~1999; Cole, Smecker-Hane \&
Gallagher~2000). 

\begin{figure}
\psfig{file=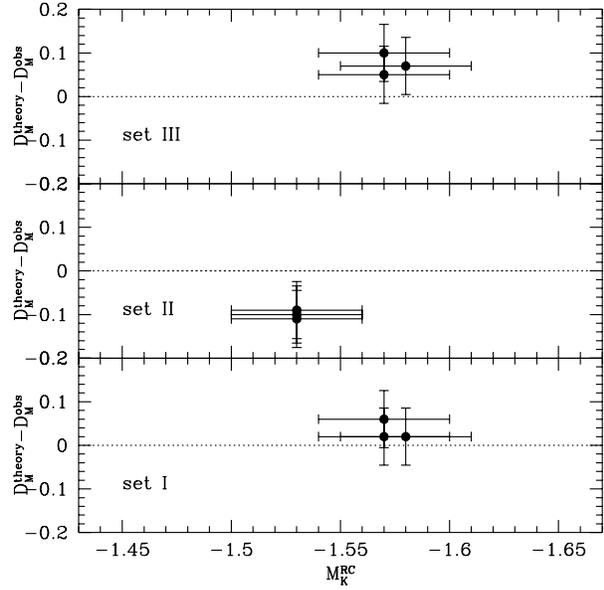,width=8.3cm}
\caption{Difference between the observed LMC $D_M$ value and
the results from the sets of models shown in Table~1, as a function
of the corresponding $M_{K}^{\rm RC}$ (see text for details).}
\label{photconstr}
\end{figure} 

Figure~\ref{photconstr} displays the comparison between observed
and predicted $D_M$ for the three sets of models discussed in the
previous section (the predicted $I$ and $K$ absolute magnitudes have
been obtained by subtracting from the observed local RC absolute
magnitude the appropriate theoretical population corrections 
summarized in Table~1); 
the error bar includes (by adding them in quadrature) 
the errors reported 
in Table~1, plus the contribution due the random errors on the
observed RC brightness in $K$ and $I$ (which are of the order of
$\pm$0.01 mag -- see, respectively, 
Pietrzy\'nski et al.~2003 and Udalski~2000),
the zero point uncertainty of 0.02 mag on the adopted reddening
maps (Udalski et al.~1999), the zero point error 
on the $K$ ($\pm$0.03 mag according to Pietrzy\'nski \& Gieren~2002)
and $I$ ($\pm$0.02 mag, Udalski et al.~1999) magnitudes, and the 0.01 error on
the photometric transformation to the $K$ Bessel \& Brett system
(see Pietrzy\'nski \& Gieren~2002).

It is interesting to notice that, at first glance, the three different
determinations of the LMC AMR 
produce models which nicely bracket the observed value of 
$D_M$, confirming our hypothesis that they constitute
realistic extreme cases for the LMC true AMR. 
The models belonging to set~I reproduce better the empirical 
constraints, whereas set~II provides values too low (by slightly more
than the 1$\sigma$ uncertainty), and set~III values slightly too high.
It goes without saying that, in case of
set~II and III results, the distances obtained from, respectively, the $K$ 
and $I$ RC magnitudes, will therefore differ at the level of about 1$\sigma$ or 
slightly more. 

Again, the AMR plays the major role in determining the result of this 
comparison. The choice of the SFR has only a minor impact, as can be
deduced from the fact that the points belonging to the same set of
models are very close to each other. Within each set of models the H99
SFR (models I$a$, II$a$ and III$a$) produces the lower theoretical 
$D_M$ value, whereas  a constant SFR throughout the galaxy lifetime 
gives origin to the higher $D_M$.
It is clear that models I$a$ (our assumed reference model in this
paper and in our previous papers) and I$c$ provide the best match of
the empirical data.

Another possible constraint to combine with the previous
one, is the magnitude 
difference between $V$ and $K$ or ($V$ and $I$), 
due to the fact that $V$ is on average
the most sensitive passband to metallicity and age (hence to SFR and
AMR) for RC stars.
However, very recently Alves et al.~(2002) found a large discrepancy (of
the order of 0.10 mag) between their $V$ magnitudes and the OGLE-II results 
for LMC stars, whereas their $K$ magnitudes are in agreement
with those of Pietrzy\'nski \& Gieren~(2002), and also their 
$HST$ $I$-band data
are consistent within the OGLE-II ones within the $HST$ photometric
zero point uncertainty of 0.02 mag.
In light of this uncertainty, 
we preferred at this stage not to use any constraint
coming from $V$-band RC photometry.

\section{Comparison with chemical abundance determinations}
\label{sec_confmet}

\begin{figure}
\psfig{file=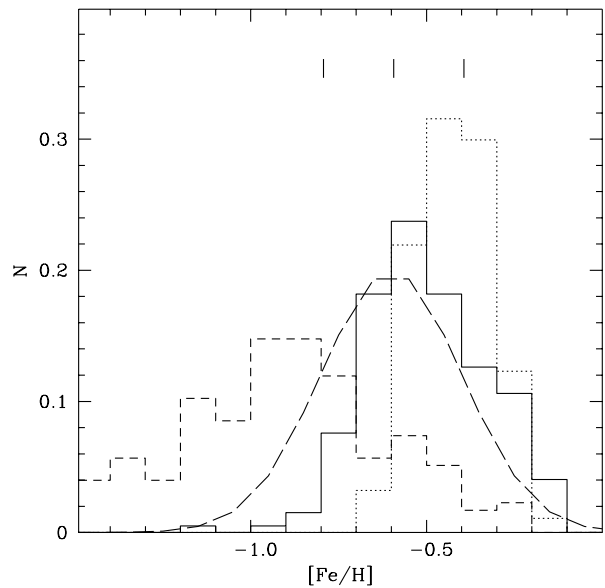,width=8.3cm}
\caption{Histogram of the number of red giants brighter than the RC as
a function of [Fe/H]. The Gaussian function (arbitrarily normalized)
reproduces the empirical data; the solid histogram refers to the results
from model I$a$, the dashed histogram denotes model I$b$, and the dotted
one denotes model I$c$. The total number of stars in
each model (about 200) has been normalized to 1.
The short lines at the top of the diagram mark the
average observed value and its upper and lower limit, according to the
systematic error quoted by Cole et al.~(2000).}
\label{Histmetgiants}
\end{figure}

Cole et al.~(2000) published results of
metallicity determinations for a sample of red giants (including asymptotic
giant branch stars above the tip of the red giant branch) brighter than
the RC in an inner disk field of the LMC. They obtained photometric
metallicities (from the Str\"omgren $m_1 - (b-y)$ diagram, which is
insensitive to stellar ages, coupled to the calibration by Hilker~2000)
for a large sample of giants, recalibrated onto Ca~II IR triplet
spectroscopic measurements of a subsample of 51 objects, tied to the
Carretta \& Gratton~(1997) spectroscopic metallicity scale for
globular clusters.
In Fig.~\ref{Histmetgiants} the best fit Gaussian of the derived
metallicity distribution is showed (arbitrarily normalized); this has
a mean value of [Fe/H]=$-$0.60, and a 1$\sigma$ dispersion of 0.20 dex.
Systematic errors at the
0.1--0.2 dex level are possible, according to the authors.
Mean metallicities of LMC field red giants which are consistent
with the Cole et al.~(2000) values, 
within the systematic uncertainty on the zero point,
have been obtained by Larsen, Clausen \&
Storm~(2000) from Str\"omgren photometry, and by
Bica et al.~(1998) from the Washington $CT_1$ system.

The metallicity distribution by Cole et al.~(2000) is compared in 
Fig.~\ref{Histmetgiants} with the results from the red giants in 
models I$a$, II$a$ and III$a$; a different choice of the SFR for a
given selected AMR does not appreciably change the [Fe/H] distribution
of bright red giants, as was the case for RC stars.
 
As for the comparison of the photometric $D_M$ discussed in the
previous section, the three different determinations of the AMR 
produce a metallicity distribution of bright red giants that nicely
brackets the observed one.
The distribution from our reference model I$a$ resembles 
the observed one; the mean [Fe/H] is [Fe/H]$\sim -$0.50, well within
the uncertainty on the Cole et al.~(2000) results.
Small variations of the dispersion around the assumed AMR may improve
the comparison, but they do not alter at all the
predicted RC levels. 

The DP97 AMR (model I$b$) provides an average
[Fe/H]$\sim -$0.40, still within the uncertainty associated with the
empirical result, whereas metallicities obtained with the
the DI00 AMR yield a mean value [Fe/H]$\sim -$0.90,
a bit too low, even considering the observational systematic errors.
This result is very similar to what was found from the photometric
comparison, where the results with the DI00 AMR where also discrepant by
slightly more than 1$\sigma$.

It is also very important to compare, before concluding this section, 
the estimate of LMC red giant metallicities obtained from
spectroscopy and from the observed red giant CMD location. 
Authors working on the tip of the red giant
branch distance scale 
(e.g. Lee, Freedman \& Madore~1993; Salaris \& Cassisi~1998)
use the $(V-I)$ colour of the red giant branch to estimate the mean
metallicity of red giants in the population under scrutiny; 
a metallicity is assigned to the
observed dereddened colour of the red giant branch 
by using relationships calibrated on Galactic globular clusters with
direct spectroscopic metallicity determinations.
In the case of the LMC, [Fe/H] values between $\sim -1.0$ and $\sim -1.3$ dex 
are found (e.g. Lee et al.~1993; Salaris \& Cassisi~1998; Sakai, Zaritski \& 
Kennicutt~2000). These metallicities are lower by about 0.4 dex than
the Cole et al.~(2000) result.

\begin{figure}
\psfig{file=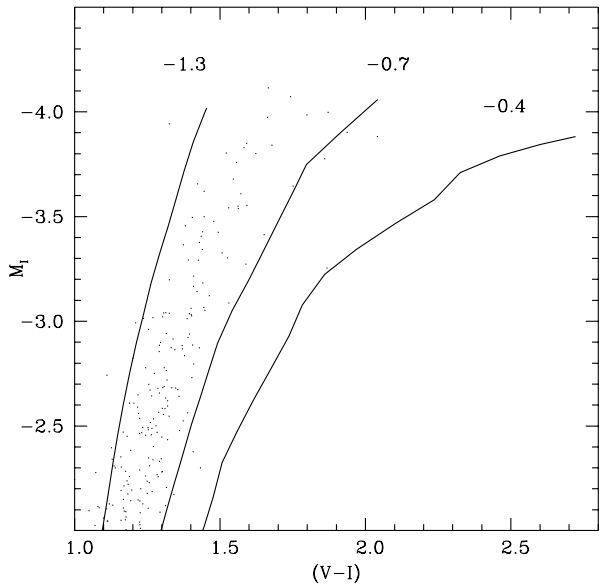,width=8.3cm}
\caption{CMD of the LMC red giant branch obtained from model~Ia (dots),
together with three 12.5 Gyr old isochrones with the labelled [Fe/H]
values.}
\label{RGBcmd}
\end{figure} 

This discrepancy is however only apparent; the reason is that Galactic
globular cluster stars have ages of the order of 12 Gyr (e.g. Salaris \&
Weiss~2002), while our simulations show that 
bright red giants in the LMC (the ones observed by
Cole et al.~2000) have average ages of the order of 1.7--1.9 Gyr .
These much younger ages shift the location of the red giants to the
blue with respect to the globular cluster counterpart at the same [Fe/H], 
thus causing an underestimate of the LMC red giants' 
metallicity when deduced from their CMD colour\footnote{see also
Davidge~(2003) for a similar conclusion in case of red giants in the
disk of NGC~6822.}.

This explanation is clearly supported
by Fig.~\ref{RGBcmd}, where we show the LMC red giant location as
deduced from our reference model (we display, for the sake of clarity,
only red giant stars up to the red giant branch tip), and the location
of 12.5 Gyr old isochrones (from the same Girardi et al.~2000 models
used in our simulations) of various metallicities.
It is evident that, when using the colour of the red giants compared
to the globular cluster counterpart, we would obtain a [Fe/H] value of
the order of $-$1.0, whereas, due to the age effect discussed above, 
the real average [Fe/H] is about $-$0.60. 

This difference between the age of LMC stars at the tip of the red
giant and the age of their Galactic globular cluster counterpart
can have an impact on their use in determining the LMC distance; we will
fully address this topic in a forthcoming paper.

\section{Discussion and conclusions}
\label{sec_concl}

In the previous sections we have shown how multiband photometry of RC
stars and metallicity determinations of red giants 
consistently indicate that the three different AMRs used to compute
our three sets of models bracket the actual LMC AMR, and that the
PT98 AMR is its best representation. Models obtained with the DI00 and
DP97 AMR constitute conservative lower and upper limits to
the LMC AMR.

We have also discussed how, in general, a constant SFR, 
or a SFR not decreasing with
time, are necessary to reproduce the RC morphology; within this
constraint, once the AMR is fixed, the difference $D_M$ between the RC
$I$ and $K$ magnitude is very weakly dependent on the SFR, albeit the
individual $I$ and $K$ values may change slightly. 
The metallicity distribution of RC stars and of red giants
is not greatly affected by the adopted SFR, but it is, of course,
very sensitive to the AMR. The average age of RC stars is typically
$\sim$1.5 Gyr, whereas the average red giant age is about 1.7--1.9 Gyr.
Due to this relatively low age of the red giants, metallicities
derived from the dereddened colour of the LMC red giant branch are
subject to a bias towards values which are too low.

To gauge the possible range of values for the LMC population
corrections, and therefore for the absolute RC
brightness in the LMC, we can consider the combinations of SFR and
AMR allowed by our sets of models which produce the brightest
and dimmest RC. The errors on the $D_M$ values (see Fig.~\ref{photconstr})
are 1$\sigma$ errors, and the error on the zero point of the RGB
metallicities (from Cole et al. 2000) we interpret, to be conservative, 
as a 1$\sigma$ error.
The reference model I$a$ fits reasonably well the central values for 
the observed 
$D_M$ and mean [Fe/H] of the red giants, therefore the difference
between the brightness of the reference model with respect to the
brightest and dimmest combination of SFR and AMR provides
a conservative estimate of the 
1$\sigma$ error bar on the theoretical population corrections.

In general, a single
combination of SFR and AMR cannot produce the brightest (dimmest)  RC in
all three photometric bands, due to the opposite behaviour of the RC
in the $K$-band with respect to the $I$ and $V$ ones.
The range of values of the RC absolute magnitude (as determined by 
subtracting the value of the appropriate theoretical population correction
from the local observed RC absolute magnitudes ) is easily estimated
from the data in table~1; the brightest RC in $V$ and $I$ is found in
model II$a$, whereas the brightest RC in $K$ is provided by models
I$c$ and III$c$.
Model III$b$ shows the dimmest RC in $V$ and $I$, and the dimmest RC
in $K$ is obtained from all three models of set~II (DI00 AMR).


The last step of our analysis is to provide a final estimate of the
LMC distance with the corresponding random and sytematic errors.
Our `best' estimate will be obtained from the $K$-band RC brightness of the 
reference model, which is the passband least sensitive
to reddening uncertainties and also shows overall the smallest systematic 
uncertainty (see Table~1). By employing the dereddened $K$
magnitude from Pietrzy\'nski \& Gieren~(2002) we obtain a distance
modulus and associated 1$\sigma$ error bar (which 
is determined by adding in quadrature the errors on
photometry, reddening and calibration discussed in 
Sect.~\ref{sec_confphot}, plus an error 
on the population corrections of $^{+0.01}_{-0.04}$ mag)
$(m-M)_{0, LMC}=18.47\pm0.01(random)^{+0.05}_{-0.06}(systematic)$
(a +0.01 mag geometric correction to the
LMC barycentre as discussed in Pietrzy\'nski \& Gieren~2002 has been 
also applied).
The distance modulus obtained from the corresponding 
$I$-band data using only the $I$-band calibration is
$(m-M)_{0, LMC}=18.44\pm0.01(random)\pm0.09(systematic)$, in
agreement with the $K$ result, but with a larger error bar (we
considered the same sources of error as for the $K$-band case, with the
exception of the error deriving from the transformations between
different $K$ bands).

We have also employed, as a consistency check, the multiband method 
applied by Alves et al.~(2002) to derive both distance and reddening
to an inner disk LMC field not covered by Udalski et al.~(1999)
reddening maps. The simultaneous determination of the apparent
distance moduli in the three photometric bands allows one 
to estimate reddening and distance at the same time, by
imposing that all three apparent distance moduli must provide the same
unreddened distance. From Alves et al.~(2002) 
data and our reference model population corrections we obtain
$(m-M)_{0, LMC}=18.49\pm0.09(random)^{+0.02}_{-0.05}(systematic)$
(we applied a correction of $-$0.013 mag for the distance to the LMC
barycentre, as in Alves et al.~2002) in beautiful
agreement with the result obtained from the dereddened $K$-band data 
by Pietrzy\'nski et al.~(2003), albeit with a somewhat larger error bar. 
The random error is the error we
obtain from the multiband fitting procedure using the absolute
magnitudes obtained from our models, including the error on the
observed RC magnitude (random and photometric zero point)
and the error on the local RC absolute brightness; 
the systematic error has been derived
by taking into account the maximum and minimum distance 
obtained by applying this method and the model combinations
displayed in Table~1.
What we have termed as random error contains in this case also the
systematic error due to the calibration of the local RC and the photometric
zero point. Due to the fact that 
the local RC brightness and the photometric zero point error 
in all of the three bands are used
simultaneously to determine both reddening and distance, it is
difficult to disentangle their systematic effect on the distance
only, and their errors -- if included in the systematic budget -- 
definetely cannot be added in quadrature to the
systematic error due to the population corrections.

The derived average reddening to the observed field is 
$E(B-V)=0.08\pm0.03(random)^{+0.06}_{-0.04}(systematic)$.
This value is consistent with the Galactic foreground reddening
in this direction, $E(B-V)=0.06\pm$0.02 (see Alves et al.~2002).
We have also checked the consistency between the reddening obtained
with this method and the reddening maps by Udalski et al.~(1999). The
idea, based on the Udalski et al.~(1999) procedure, is to determine the
difference between the observed $I$ magnitude of Alves et al.~(2002)
and the corresponding quantity provided by   
the OGLE-II fields I and II with known reddening (E(B-V)=0.152).
Once the Schlegel et al.~(1998) extinction law is
adopted, and an intrinsic difference by 0.02 mag due to geometrical effects 
is accounted for, we can therefore
determine the reddening of Alves et al~(2002) field 
on the same scale as the OGLE-II maps.  
From this procedure we obtain $E(B-V)=0.11\pm$0.02, in agreement,
within the error bars, with our previous result.

To conclude, we have provided a best estimate of the LMC distance modulus 
from the RC method, 
$(m-M)_{0,LMC}=18.47\pm0.01(random)^{+0.05}_{-0.06}(systematic)$,
where the systematic error has been realistically and carefully 
determined, based on the actual photometric and spectroscopic observations
of the LMC stellar population. The size of this systematic error is
small and highlights the fact that, when observations are able to 
constrain the SFR and AMR of the population under scrutiny, and
provided the population corrections are appropriately computed as
discussed in our series of papers, the RC method 
(especially in the $K$-band, in the case of the LMC) 
can provide accurate distances.

\section*{Acknowledgments}

SMP acknowledges financial support from PPARC.


\label{lastpage}

\end{document}